\documentclass[twocolumn,prb,superscriptaddress,floatfix]{revtex4}

\usepackage{graphicx}

\usepackage{dcolumn}

\newcommand{\ctd}{\ensuremath{C_{\text{2DEG}}}}
\newcommand{\co}{\ensuremath{C_{1}}}
\newcommand{\ct}{\ensuremath{C_{2}}}
\newcommand{\csig}{\ensuremath{C_{\Sigma}}}
\newcommand{\cg}{\ensuremath{C_{g}}}
\newcommand{\cs}{\ensuremath{C_{s}}}
\newcommand{\ro}{\ensuremath{R_{1}}}
\newcommand{\rt}{\ensuremath{R_{2}}}
\newcommand{\ec}{\ensuremath{E_{c}}}
\newcommand{\qo}{\ensuremath{Q_{0}}}

\newcommand{\te}{\ensuremath{T_{e}}}
\newcommand{\units}[1]{\ensuremath{\mathrm{#1}}}
\newcommand{\amount}[2]{\ensuremath{#1~\units{#2}}}

\begin{document}

\DeclareGraphicsExtensions{.eps, .jpg}

\title{Charge Transport Processes in a Superconducting Single-Electron 
Transistor Coupled to a Microstrip Transmission Line}
\author{W. Lu}
\affiliation{Department of Physics and Astronomy, Rice University,
Houston, Texas 77005}
\author{K. D. Maranowski}
\affiliation{Materials Department, University of California, Santa Barbara, 
California 93106 }
\altaffiliation[Current address: ]{Cielo Communications, Inc., 325
Interlocken Parkway, Broomfield, CO 80021.}
\author{A. J. Rimberg}
\affiliation{Department of Physics and Astronomy, Rice University,
Houston, Texas 77005}
\affiliation{Department of Electrical and Computer Engineering, Rice University, 
Houston, Texas 77005}

\begin{abstract}
We have investigated charge transport processes in a superconducting
single-electron transistor (S-SET) fabricated in close proximity to a
two-dimensional electron gas (2DEG) in a GaAs/AlGaAs heterostructure. 
The macroscopic bonding pads of the S-SET along with the 2DEG
form a microstrip transmission line.   We observe a variety of
current-carrying cycles in the S-SET which we attribute to
simultaneous tunneling of Cooper pairs and emission of photons into the
microstrip.  We find good agreement between these experimental
results and simulations including both photon
emission and photon-assisted tunneling due to the electromagnetic
environment.
\end{abstract}

\maketitle

Charge transport in all-superconducting single-electron transistors
(S-SETs) has been studied in detail, and numerous current-carrying
cycles have been identified, including the Josephson-quasiparticle (JQP)
cycle,\cite{Fulton:1989,vandenBrink:1991a,Nakamura:1996,Nakamura:1997}
the 3$e$ cycle\cite{Hadley:1998} and Andreev
reflection.\cite{Fitzgerald:1998b}  What processes occur in a given
S-SET is determined by a complex interplay between Cooper pair tunneling
and charging effects, leading to rich behavior in  S-SET current-voltage
characteristics.  Further complexity is added by the electromagnetic
environment, which can cause photon-assisted tunneling (PAT) of
quasiparticles\cite{Martinis:1993,Hergenrother:1994} and incoherent
tunneling of Cooper pairs.\cite{vandenBrink:1991,Siewert:1996}
Experiments have addressed the influence of the environment on a variety
of tunnel junction systems, including electron
pumps,\cite{Martinis:1993,Covington:2000} single tunnel
junctions,\cite{Holst:1994} Josephson junction arrays\cite{Rimberg:1997}
and S-SETs fabricated above a ground plane, focusing on the
supercurrent.\cite{Kycia:2001}  The effect of the environment on S-SETs
remains interesting given the potential use of Coulomb blockade devices
for quantum computation.\cite{Nakamura:1999} Here we report measurements
on samples similar to that in Ref.~\onlinecite{Kycia:2001}, but focus on
transport at higher biases. We observe previously unreported charge
transport cycles which we attribute to Cooper pair tunneling with
simultaneous emission of a photon into the environment. We find good
agreement between our measurements and simulations based on the orthodox
theory of electron tunneling,\cite{Likharev:1988,Grabert:1992} including
the effects of both photon emission and PAT\@.

\begin{figure}[b]
\includegraphics{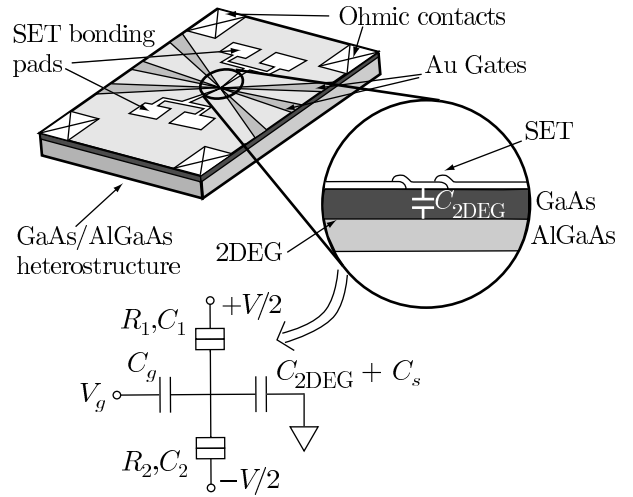}
\caption{\label{sample}Schematic
diagram of a sample, showing the macroscopic bonding pads and Au gates. 
Expanded view: cross-section of the tunnel junctions and the
capacitance \ctd\ coupling the SET to the 2DEG\@.  Circuit diagram:
specific parameters discussed in the text, including the
parallel combination of \cs\ and \ctd.}
\end{figure}

Our samples consist of Al/AlO$_{x}$/Al S-SETs fabricated on a
GaAs/AlGaAs heterostructure containing a two-dimensional electron gas
(2DEG) located  $\sim\amount{50}{nm}$ below the substrate surface, as
illustrated in Fig.~\ref{sample}. The heterostructure, grown on a GaAs
substrate using molecular beam epitaxy, consists of the following
layers: \amount{1000}{nm} of GaAs, \amount{47}{nm} of
Al$_{0.3}$Ga$_{0.7}$As and \amount{5}{nm} of GaAs.  The
Al$_{0.3}$Ga$_{0.7}$As is delta-doped with Si \amount{22}{nm} from the
lower GaAs/Al$_{0.3}$Ga$_{0.7}$As interface, at which the 2DEG forms. 
The SETs are fabricated using electron-beam lithography and shadow
evaporation techniques. The central island of the SET is coupled by a
capacitance \ctd\ to the 2DEG beneath it, which is held at ground. The
SET is surrounded by six Au gates, as described elsewhere.\cite{Lu:2000}

We model the SET as shown, with tunnel junction resistances and
capacitances $R_{1(2)}$ and $C_{1(2)}$ respectively. A voltage $V_{g}$
applied to one of the Au gates is used to adjust the electrostatic
potential of the SET island through the capacitance \cg. The remaining
gates are held at ground and contribute \cs\ to the total island
capacitance $\csig = \co +\ct +\cg + \cs +\ctd$.  We focus first on the
2DEG far from the SET, which in conjunction with the macroscopic bonding
pads forms a microstrip transmission line to which the SET is coupled. 
Such microstrip can resonate at microwave frequencies and thereby affect
transport in tunnel junction systems.\cite{Holst:1994,Kycia:2001} We
have examined two such samples with junction resistances ranging from
\amount{24}{k\Omega} to \amount{192}{k\Omega}, and a third control
sample fabricated on a piece of GaAs without a 2DEG. Sample parameters
are given in Table~\ref{params} below. 

\begin{figure}[t!]
\includegraphics{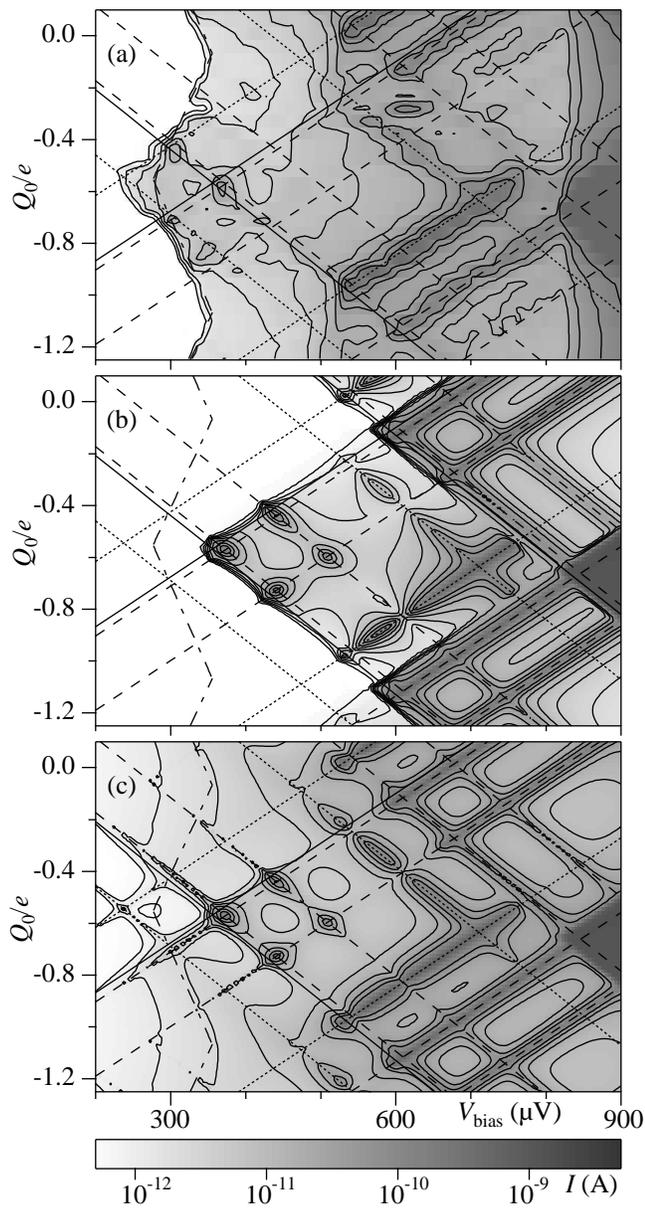}
\caption{\label{images} (a) Measured $I$ for Sample 1 on a logarithmic
scale  vs.\ $V$ and \qo.  (b) Simulated $I$ using the parameters in
Table~\ref{params}, without PAT\@.  (c) Simulated $I$ including PAT,
using $R_{e} = \amount{9}{\Omega}$ and $\te = \amount{3}{K}$. Lines
correspond to quasiparticle (solid), Cooper pair (dotted), Cooper pair
with photon emission (dashed) and 3$e$ (dash-dotted) transitions.}
\end{figure}

The large value of $\ctd\approx\amount{356}{aF}$ (\amount{382}{aF}) for
Sample 1 (2), along with the total gate capacitance
$\cs+\cg\approx\amount{20}{aF}$ helped ensure that the charging energy
$\ec=e^{2}/2\csig$ satisfied $\ec < \Delta$ where $\Delta$ is the
superconducting gap.  For the control sample, \co\ and \ct\ were 
sufficiently large to give a comparable \ec. For all the samples, the
Josephson energy given by the Ambegaokar-Baratoff
relation\cite{Ambegaokar:1963} satisfied $E_{J_{i}}=h\Delta/8e^{2}R_{i}
< \ec$ so that the charge state $n$ of the island is well defined,
justifying use of the orthodox theory.  We define the charging energy of
the SET $U(n) = \ec(n - Q_{0}/e)^2$, where $Q_{0}=\cg V_{g}$ is the
offset charge induced by the gate voltage.  At low temperatures,
electrons will tunnel through a given junction only when the change in
free energy $\delta f = f_{f} - f_{i}>0$, including both changes in
$U(n)$ and work performed by the bias voltage $V$. For $V<4\Delta$
single-electron tunneling is in general forbidden by the combined
effects of the Coulomb blockade and the superconducting gap, though more
complex charge transport processes such as the JQP and 3$e$ cycles are
allowed at specific locations in the \qo-$V$ plane.

We measured the samples in a dilution refrigerator at
$T_{\text{mix}}=\amount{20}{mK}$ using a four probe configuration
symmetrically biased with respect to ground. High-frequency noise was
excluded using standard techniques as described elsewhere.\cite{Lu:2000}
Sample parameters such as the capacitances and summed junction
resistances $\ro+\rt$ were determined from the measurements. The
resistance ratio $\ro/\rt$ was chosen to give the correct relative size
of features associated with each junction in the simulations.  Current
$I$ was measured versus bias voltage $V$ for a series of different
values of the offset charge \qo\ for all three samples; results for
Sample 1 are presented in Fig.~\ref{images}(a).

\begin{table}
\caption{\label{params}Physical parameters of the samples. Energies are
in $\mu$eV, capacitances in aF and resistances in k$\Omega$.}
\begin{ruledtabular}
\begin{tabular}{cccccdd}
Sample & $\Delta$  & \co  & \ct & \ec% 
& \multicolumn{1}{c}{$R_{1}+R_{2}$} & \multicolumn{1}{c}{$R_{1}/R_{2}$} \\
1 & 207 & 181 & 120 & 118 & 305 & 0.59 \\
2 & 201 & 375 & 260  & 77 & 63.5 & 0.6 \\
control & 198 & 515 & 322  & 91 & 91. & 0.7 \\
\end{tabular}
\end{ruledtabular}
\end{table}

\begin{figure}
\includegraphics{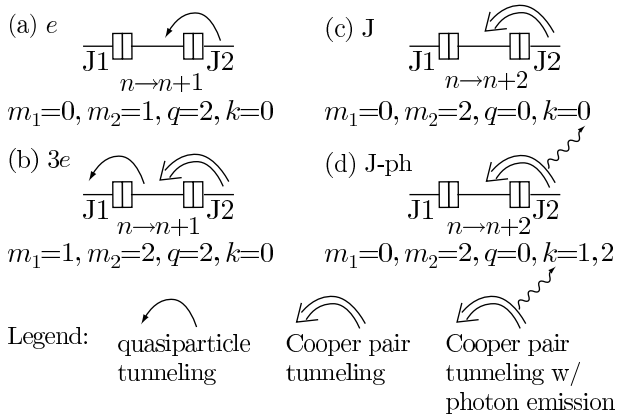}
\caption{\label{processes} Tunneling processes discussed in the text,
 for positive bias voltage and increasing island charge.}
\end{figure}

Several resonances are visible for voltages between 500 and
\amount{800}{\mu V}. The most pronounced are the usual JQP peaks, some
of which are paralleled by two smaller resonances.  We believe these
additional resonances are analogs of the usual JQP peaks, but differ in
that Cooper pairs emit one or more photons into the transmission line
while tunneling. We refer to them as JQP-ph peaks. Several smaller
features are also visible near $\qo=0.5$ at biases between 300 and
\amount{450}{\mu V}.  To support our interpretation of the JQP-ph
resonances, and to identify the smaller features, we begin with
conservation of energy, including not only the usual terms accounting
for creation of quasiparticles and changes in charging
energy,\cite{Fitzgerald:1998b,Pohlen:1999a} but also
a term allowing for emission of photons:
\begin{eqnarray}
{\textstyle\sum_{i}\kappa_{i}m_{i}}eV & = & U(n+\delta m)-U(n) +q\Delta +%
 k\hbar\omega_{s} \\
& = & 2\delta m\ec\left [-(\frac{\qo}{e} - n) + \frac{\delta m}{2}\right ] %
\nonumber \\
& & + q\Delta + k\hbar\omega_{s} \nonumber 
\end{eqnarray}
Here $\kappa_{i} = 1 - (C_{i}+\cg)/2$ is the fraction of the bias
voltage available across junction $i$ and $m_{i}$ the number of
electrons transferred across it, $\delta m = m_{2}-m_{1}$ is the change
in the island charge, $q$ is the number of quasiparticles created, and
$k$ is the number of photons of energy $\hbar\omega_{s}$ emitted.  The
proper choice of $m_{1}$, $m_{2}$, $q$ and $k$, gives transition rules
for quasiparticle, Cooper pair, Cooper pair with photon emission, and
3$e$ tunneling processes. These are indicated schematically in
Fig.~\ref{processes}, and transition lines corresponding to them are
shown in Fig.~\ref{images}.

The JQP features in Fig.~\ref{images}(a) lie along Cooper pair
transition lines, either with (J-ph) or without (J) photon emission,
since Cooper pair tunneling is a resonant process. For the primary ($k=0$) JQP
peak between $\qo\approx -1.0$ and $-0.6$, the cycle begins with a
Cooper pair transition through J1 ($-1 \rightarrow -3$), followed by
successive quasiparticle transitions through J2 ($-3 \rightarrow
-2 $ and $-2 \rightarrow -1$) which return the island to its original
charge state and allow the cycle to begin again.  As it must, the JQP
peak begins just above the quasiparticle line for the $-2 \rightarrow -1
$ transition, which gives the requirement $eV>(\ec+2\Delta)\approx
\amount{533}{\mu eV}$.

To address the JQP-ph peaks, the fundamental frequency for photon
emission was determined by adjusting $\hbar\omega_{s}$ so that a
one-photon J-ph line coincided with the first JQP-ph resonance between
$\qo = -1.2$ and $-0.8$. Best results were obtained for
$\hbar\omega_{s}=\amount{136}{\mu eV}$. For Cooper pair tunneling with
$k=1$ we obtain the condition
$eV>(\ec+2\Delta+\hbar\omega_{s}/2)\approx\amount{601}{\mu eV}$, in good
agreement with the observed location of the resonance. Since the typical
lead width ($>\amount{20}{\mu m}$) for the pad structures is much
greater than the 2DEG depth of \amount{50}{nm}, the effective wavelength
for excitations on the microstrip\cite{Pozar:1998} is
$\lambda_{e}\approx 2\pi
c/\sqrt{\epsilon}\omega_{s}\approx\amount{2.5}{mm}$ where
$\epsilon\approx 13$ is the dielectric constant of GaAs.  The
fundamental resonance corresponds to a length
$\ell=\lambda_{e}/2\approx\amount{1.25}{mm}$, in agreement with the
known dimensions of our lead structures. Resonances with a best fit for
$\hbar\omega_{s}=\amount{160}{\mu eV}$ were observed in Sample 2, while
none were observed in the control sample, supporting our interpretation
of the data. The position of the $k=2$ J-ph line does not match well to that of
the associated resonance, perhaps not surprising given its small size and
the large background current.

To gain further insight into the features in Fig.~\ref{sample}(a),
we turn to our simulations. Here, to account for the JQP-ph resonances,
we make a phenomenological modification of the theory of Averin and
Aleshkin\cite{Averin:1989a,Aleshkin:1990} and write for the Cooper pair
tunneling rate $\Gamma_{\text{cp}}$
\begin{equation}
\Gamma_{\text{cp}}=\sum_{k=0}^{2}A_{k}\frac{\Gamma_{\text{qp}}E_{J}^{2}}%
{4(\delta f -k\hbar\omega_{s})^{2}+(\hbar\Gamma_{\text{qp}})^{2}}
\end{equation}
where $\delta f$ is the change in free energy due to the tunneling of
the Cooper pair and $\Gamma_{\text{qp}}$ is the rate of the first
subsequent quasiparticle tunneling.  The $A_{k}$ were chosen as
$A_{0}=1$ (giving the usual Averin-Aleshkin term) and $A_{k} =
\frac{1}{k}(\ec/\hbar\omega_{s})^{k}\exp(-\ec/\hbar\omega_{s})$, the
weights\cite{Grabert:1992} for independent emission of photons into a
single mode.\footnote{While this approach gives reasonable agreement
with the measured peak heights, the JQP-ph features may in fact be due
to harmonics of the stripline resonance. None of our main conclusions
are affected, however, by whether the higher J-ph resonances are due to
single or multiple photons.}

Results of a simulation including the effects of photon emission are
shown in Fig.~\ref{images}(b).  The agreement is reasonable, especially
at higher biases, where the $k=1$ and 2 JQP-ph lines appear.
Moreover, the simulation predicts the appearance of a feature at the
intersection of the $k=1$ J-ph lines for J1 and J2 near $\qo=-0.58$ and
$V \approx\amount{374}{\mu V}$. A corresponding peak can also be seen in
the data in Fig.~\ref{images}(a).  This feature is an analog of the $3e$
peak\cite{Hadley:1998} but with Cooper pair tunneling replaced by J-ph
processes, and we refer to it as the 3$e$-ph peak.  Our samples do not
satisfy the condition $\ec>\frac{2}{3}\Delta$ for the appearance of the
usual 3$e$ cycle.  The corresponding condition for the 3$e$-ph cycle in
Sample 1 is $\ec>\frac{2}{3}(\Delta -
\hbar\omega_{s}/4)=\amount{115.6}{\mu eV}$ which is just satisfied.  A
similar feature appears in the data (at $\qo\approx-0.73$ and
$V\approx\amount{440}{\mu V}$) near the intersection of a $k=1$ J-ph
line for J1 and a $k=2$ J-ph line for J2. A corresponding peak appears
at the same location in the simulation.

Despite this level of agreement, some discrepancies exist.  The measured
current remains substantial down to the 3$e$ threshold line between
275--$\amount{350}{\mu V}$, while the calculated current in
Fig.~\ref{images}(b) nearly vanishes for all \qo\ below the 3$e$-ph
peak.  Furthermore, in the simulation the primary ($k=0$) JQP peak is
strongly suppressed at the intersections with J-ph transition lines; an
example occurs in Fig.~\ref{images}(b) at $\qo\approx-0.83$ and $V=
\amount{607}{\mu V}$ where the primary JQP peak for J1 intersects a
$k=2$ line for J2.  The standard JQP process would begin with the Cooper
pair transition $-1\rightarrow-3$. Our simulations indicate that at this
intersection the J-ph transition $-1\rightarrow 1$ is allowed, and is
followed rapidly by a $1\rightarrow 0$ quasiparticle transition.  The
transition $0\rightarrow -1$ is allowed but slow, so that the
occupational probability of the $n=-1$ state is quite small ($\sim
0.01$), and the JQP cycle is suppressed.  There is, however, no clear
sign of suppression of the primary JQP cycle in our data.
\begin{figure}[ht!]
\includegraphics{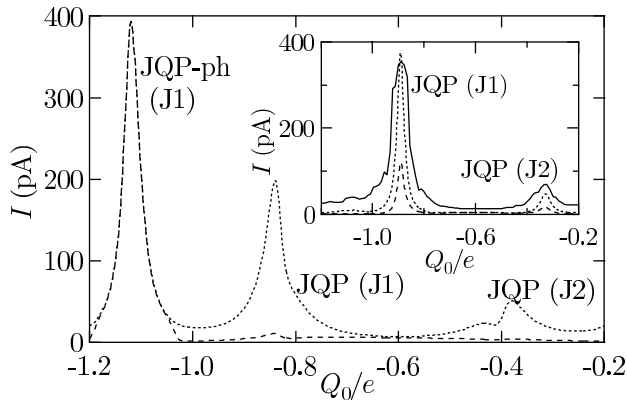}
\caption{\label{jqp}Simulated $I$ vs.\ \qo\ for $V=\amount{607}{\mu V}$,
with (dotted line) and without (dashed line) the effects of PAT\@. 
Inset: measured $I$ vs.\ \qo\ for Sample 1 at $V \approx\amount{577}{\mu
V}$ (solid), and simulated $I$ vs.\ \qo\ with (dotted) and without
(dashed) PAT\@.  Here $\te=\amount{2}{K}$ and $R_{e}=\amount{3}{\Omega}$.}
\end{figure}

We can account for these discrepancies to some extent by considering PAT
due to the environment, here assumed to be the 2DEG in the immediate
vicinity of the SET\@.  Since the effective environment temperature \te\
can be much larger than the SET electron
temperature,\cite{Martinis:1993} PAT can be quite significant. Following
Siewert and Sch\"{o}n,\cite{Siewert:1996} we incorporate it into our
calculation by adding the rate
\begin{equation}
\Gamma_{e} =%
 \frac{\pi}{e^{2}R_{i}}\frac{R_{e}}{2R_{K}}k_{B}\te\frac{2\Delta}{2\Delta +%
\delta f}\exp[-(2\Delta+\delta f)/k_{B}\te]
\end{equation}
to all quasiparticle tunneling events for which $\delta f +2\Delta>0$.
Fig.~\ref{images}(c) shows the results of a simulation including PAT
with much improved agreement.  Appreciable current appears at voltages
around the 3$e$ threshold lines.

In addition, the suppression of the JQP peak at intersections with J-ph
lines is largely lifted.  This can be seen in greater detail in
Fig.~\ref{jqp} which shows the calculated $I$ vs. \qo\ for
$V=\amount{607}{\mu V}$, both with and without PAT\@.  The inclusion of
PAT increases the primary JQP peaks for both J1 and J2.  In the inset,
we make a comparison with data from Sample 1 at $V
\approx\amount{577}{\mu V}$, a voltage below the cutoff for the JQP-ph
cycle, so that only the JPQ peak is visible.  The JQP peaks for both J1
and J2 are near J-ph transition lines here, and in the absence of PAT
the simulation underestimates the actual JQP current. Addition of PAT
brings the simulation in much closer agreement with the data, although
it does not correctly predict the width of the JQP peak, a topic which
will be addressed in a separate publication.

In conclusion, we have investigated charge transport processes in an
S-SET which is coupled to a 2DEG, so that the contact pad structures 
form microstrip resonators.  We have observed current-carrying cycles
which involve tunneling of a Cooper pair and simultaneous photon
emission, including JQP-ph and 3$e$-ph processes.  Simulations performed
using the orthodox theory agree well with the data, particularly when
the effects of PAT due to the environment are included.  In the
simulations, suppression of the primary JQP in the vicinity of J-ph
lines is partially lifted by the effects of PAT\@.  While we do not
observe the suppression, it may be detectable in samples for which the
cental island of the SET is not so strongly coupled to an external
environment.

This research was supported at Rice by the NSF under Grant No.\
DMR-9974365 and by the Robert A. Welch foundation, and at UCSB by the
QUEST NSF Science and Technology Center.  One of us (A. J. R.)
acknowledges support from the Alfred P. Sloan Foundation.

%\bibliography{TunnelJunction,QuantumCoherence,QuantumDot}

\end{document}